\documentclass{emulateapj}
\shortauthors{Wang et al.}

\newcommand{\go}{\mathrel{\raise.3ex\hbox{$>$}\mkern-14mu
             \lower0.6ex\hbox{$\sim$}}}

\begin{document}

\title{Optical I-band Linear Polarimetry of the Magnetar 4U 0142$+$61 with Subaru}

\author{Zhongxiang Wang\altaffilmark{1}, 
Yasuyuki~T. Tanaka\altaffilmark{2},
Chen Wang\altaffilmark{3},
Koji~S. Kawabata\altaffilmark{2},
Yasushi Fukazawa\altaffilmark{4},
Ryosuke Itoh\altaffilmark{4},
Anestis Tziamtzis\altaffilmark{1}
}

\altaffiltext{1}{Shanghai Astronomical Observatory, Chinese Academy of Sciences,
80 Nandan Road, Shanghai 200030, China}

\altaffiltext{2}{Hiroshima Astrophysical Science Center, Hiroshima University, 
Higashi-Hiroshima, Hiroshima 739-8526, Japan}

\altaffiltext{3}{National Astronomical Observatories, 
Chinese Academy of Sciences, A20 Datun Road, Beijing 100012, China}

\altaffiltext{4}{Department of Physical Sciences, Hiroshima University, 
Higashi-Hiroshima, Hiroshima 739-8526, Japan}

\begin{abstract}
Magnetars are known to have optical and/or infrared emission, but
the origin of the emission is not well understood. 
In order to fully study their emission properties, we have carried out
for the first time optical linear polarimetry of the magnetar 4U~0142+61, 
which has been determined from different observations to have a complicated 
broad-band spectrum over optical and infrared wavelengths. 
From our $I$-band imaging polarimetric observation, conducted with
the 8.2-m Subaru telescope, we determine the degree of linear polarization 
$P=1.0\pm$3.4\%, or $P\leq$5.6\% (90\% confidence level). 
Considering models suggested for optical emission from magnetars, 
we discuss the implications of our result.
The upper limit measurement indicates that different from radio pulsars, 
magnetars probably would not have strongly polarized optical emission if
the emission arises from their magnetosphere as suggested.
\end{abstract}


\section{INTRODUCTION}
It is currently accepted that magnetars are highly magnetized neutron 
stars whose dipole magnetic field reaches $\sim$10$^{14}$--10$^{15}$~G,
implied by their slow spin periods of 2--12\,s and rapid spin-down rates 
of $10^{-12}-10^{-10}$ s s$^{-1}$ (see reviews, e.g., given 
by \citealt{wt06,kas07,mer13}; but see \citealt{rea+10,rea+12} for
two low magnetic field magnetar cases recently discovered). 
Neutron stars which have traditionally been classified as anomalous X-ray 
pulsars (AXPs) or soft Gamma-ray repeaters (SGRs) belong to this magnetar 
class, and over 20 magnetars have been identified up to now \citep{ok14}. 
Their bright X-ray luminosities cannot be explained by rotational
energy loss rates and hence the energy source is considered to be provided by
the decay of their ultra-strong magnetic fields. AXPs and SGRs are well known 
to show a variety of high-energy phenomena \citep[e.g.,][]{wt06,kas07} 
including the exceptionally bright `giant flares' whose peak luminosity 
amounts to $\sim 10^{46}$ erg s$^{-1}$ \citep[e.g.,][]{Hurley99, Terasawa05, Tanaka07}.

In addition to these intensive studies at X-ray energies, magnetars have 
been subjected to multi-wavelength follow-up observations in radio, optical, 
and infrared (IR) bands.  Among the magnetars identified so far, 
the AXP 4U 0142+61 is the best studied magnetar at optical and IR wavelengths 
due to 
its relatively short distance (distance $d\simeq 3.6$ kpc) and low extinction 
($A_V\simeq3.5$; \citealt{dv06a,dv06b}). An optical counterpart was first 
discovered by \citet{hvk00} from 4U~0142+61. The optical flux was higher 
than the Rayleigh-Jeans tail of the blackbody component seen in X-ray band, 
requiring another emission mechanism. Subsequently, \citet{km02} found that 
the optical emission is pulsed at its spin period and that the pulsed fraction 
is $\sim$27\%, which is higher than 4\%--14\% observed in its X-ray 
emission \citep{gon+10}. A very faint near-IR counterpart was also identified 
by \citet{Hulleman04}, and then \textit{Spitzer Space Telescope} detected 
mid-IR emission from 4U~0142+61 \citep*{wck06}. 

Importantly, \citet{wck06} found that two different emission components are 
needed to explain the combined optical and IR spectral energy 
distribution (SED). Namely, the thermal blackbody-like component, possibly 
emitted from a debris disk around the pulsar, is dominant over 
the 2.2--8 $\mu$m range, while another power-law like component 
(represented by $F_{\nu} \propto \nu^{0.3}$) can account for 
the residual emission in the optical $VRI$ and near-IR $J$ bands \citep{wck06}. 
Follow-up \textit{Spitzer} mid-IR spectroscopy and 24~$\mu$m imaging of 
the AXP were consistent with the interpretation of a debris disk for 
the IR emission \citep{wck08}. However, the origin of the optical power-law 
emission is still unclear, although it would probably arise from 
the magnetosphere because the emission is relatively highly pulsed.

In order to shed new light on the optical emission mechanism of magnetars, 
we have carried out polarimetric observations for the nearest and least 
absorbed magnetar 4U~0142+61 using the Subaru 8.2-m telescope. 
In \citet*{wtz12}, we have reported a 4.3\% upper limit 
(90\% confidence level) on the degree of circular polarization in $I$-band 
emission from the source. In this paper we present the result from our linear 
polarimetry of the source at $I$-band.

\section{OBSERVATION AND DATA REDUCTION}    
\label{sec:obs}

We carried out linear imaging polarimetry of 4U 0142+61 at I-band with
the 8.2-m Subaru Telescope on 2013 December 22--23. 
Two first half of the telescope time in each night were awarded 
to us by the National Astronomical Observatory of Japan. 
Due to high humidity in the first night, only three hours of data were 
taken. For the second night, full five hours of the telescope time were 
used. 

The Faint Object Camera and Spectrograph 
(FOCAS; \citealt{kas+02}) was used for the polarimetric observation.
In this mode, a Wollaston prism and a half-wave retarder are inserted 
to the collimated beam. 
An incident beam is splited by the Wollaston prism into two orthogonally 
polarized beams, one ordinary (o-beam) and the other extraordinary (e-beam).
To avoid blending of the two beams, a standard focal mask
has to be used. In order to
measure the degree of linear polarization and the position angle 
of polarization in the celestial plane, one set of four exposures of a target
with the half-wave plate at four position angles, 
0\fdg0, 45\fdg0, 22\fdg5, and 67\fdg5, are required.
The FOCAS detector was two fully-depleted-type 2k$\times$4k CCDs, the pixel 
scale of which was 0\farcs104\,pixel$^{-1}$. We 
2$\times$2 binned the detector for our observation.

In our observation, each set of four exposures were taken by setting
the half-wave plate at the four position angles alternately. 
In order to avoid possible sever saturation caused by bright stars, 
the time of each exposure was 2 min.
Between the sets of the exposures, the telescope was five-point dithered 
to avoid bad pixels on the CCDs and to help remove cosmic ray hits during
the data reduction. 
In total, in the first and second night we obtained 19 sets and
32 sets of the exposures, respectively. 
The observing conditions were unfortunately
mediocre comparing to what we wished (0\farcs5 seeing would be ideal).
The average seeing was approximately 0\farcs80 in 
the first night and 0\farcs86 in the second night.
\begin{center}
\includegraphics[scale=0.58]{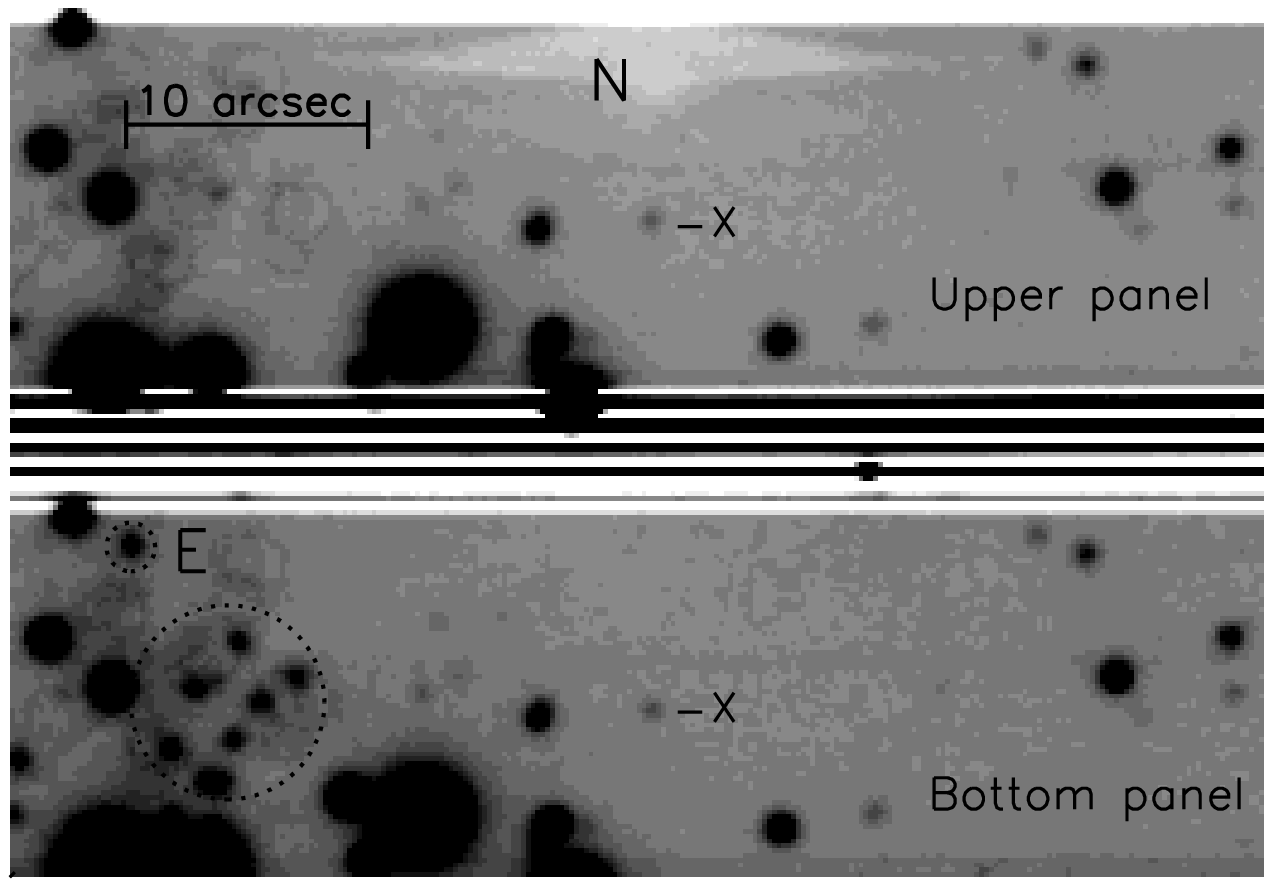}
\figcaption{Subaru/FOCAS linear polarization image of 
the 4U~0142+61 field at I-band.
The two polarized beams were recorded at the upper and bottom panels, in which 
the AXP is indicated by $X$. Several ghost stars, marked by dotted circles, 
are present in the bottom panel.
\label{fig:fld}}
\end{center}

In order to determine the zero point of the position angle of polarization
in the celestial plane, the standard star BD+64d106, which has strongly 
polarized emission, was observed.

We used the IRAF packages for data reduction. The images were bias subtracted
and flat fielded. Dome flats at the four position angles of the half-wave plate
were taken and used for flat fielding respectively. 
For our target, the images made at each 
angle were then combined into one final image of the target field
by positionally calibrating them to a reference image. In the first
night at the position angle of 45\fdg0,  the target position in one image
was contaminated by a cosmic ray hit, and the image was not included in
image combining. In total the on-source exposure times at each position
angle is 38 min (36 min for the position angle of 45\fdg0) and 64 min
in the first and second night, respectively. 
Because of the faintness of the target and mediocre seeing during our 
observations, the signal-to-noise ratios of the target
in each night's images are not sufficiently high. 
We therefore combined all the data of the two nights and made a set of four
images of the target field at the four position angles respectively.
A target-field image,
made by combing all the images, is shown in Figure~\ref{fig:fld}.

For the standard star, aperture photometry was performed and the o-beam
and e-beam brightnesses at each position angle were obtained.
For the target, because of its faintness, we used 
the point spread function (PSF) photometry tasks from IRAF's DAOPHOT package
to measure its brightnesses. 
We also derived aperture corrections by using
8\ in-field, relatively bright stars (brightest stars in the field 
were saturated).
A radius of 10.0 pixels ($\simeq 2\farcs1$)
was used for aperture photometry of these bright stars.
The uncertainties on the corrections are negligible, as they
are much smaller 
than those from photometry of the target.
These magnitude measurements and uncertainties 
are given in Table~\ref{tab:phot}.

\section{RESULTS} 
\label{sec:res}

The degree of linear polarization $P$ is calculated from
\begin{equation}
P=\sqrt{Q^2+U^2}\ ,
\end{equation}
where $Q$ and $U$ are Stokes parameters. They are determined from
\[
Q=\frac{1-a_1}{1+a_1},\ \  U=\frac{1-a_2}{1+a_2}\ ,
\]
where 
\[
a_1=\sqrt{\frac{\kappa_{0\fdg0}}{\kappa_{45\fdg0}}},\ \  a_2=\sqrt{\frac{\kappa_{22\fdg5}}{\kappa_{67\fdg5}}}\ .
\]
Here $\kappa=I_e/I_o$, the ratio of the intensity $I$ of the target in
the e-beam and o-beam frames at each of the four position angles of the
half-wave plate.  The position angle $\psi$ of linear
polarization is determined from
\begin{equation}
\psi=\frac{1}{2}\arctan(\frac{U}{Q})\ .
\end{equation}
Then from standard propagation of uncertainty,  where a first-order
Taylor series expansion is used and the variables
in the functions are assumed to be independent,
the uncertainties on $P$ and $\psi$ can be calculated from those on
intensity measurments (see, e.g., \citealt{kaw+99}).

Using the formulae given above, we first checked the polarization 
measurements for the standard star.
From the intensity measurements,  we found $P=4.710\pm0.046\%$ 
($Q=2.325\pm 0.046\%$, $U=-4.096\pm 0.046\%$)
at $I$-band,
which is consistent with the reported value 4.696$\pm0.052\%$
\citep*{sel92}. Therefore no instrumental correction to the polarization
measurement was needed.
The position angle $\psi'$ in the instrumental 
coordinate (cf. Equation (2)) was found to be 
$\psi' =-30\fdg21\pm0\fdg28$.
Comparing it to the reported value 
$\psi=96\fdg89\pm0\fdg32$ \citep{sel92}, which is the position angle in 
the celetial plane, the correction $\delta$ was 
determined to be $\delta=\psi'-\psi=-127\fdg10\pm0\fdg43$. 

We then calculated the Stokes parameters and degree of linear polarization
for 4U 0142+61, and found $Q=-0.97\pm3.40\%$, $U=-0.21\pm3.31\%$, 
and $P=1.0\pm3.4\%$. Since the uncertainty dominates, a 90\% confidence level
constraint on the degree of linear polarization was derived. The upper limit
is 5.6\%.  In addition, also due to the large uncertainties on $U$ and $Q$,
$\psi=\psi'-\delta= 43\arcdeg\pm96\arcdeg$ (where $\psi' =-83\fdg9$), 
which provides no useful information.

\section{DISCUSSION}
\label{sec:dis}

Neutron stars are known to generally have faint optical emission, either 
the Rayleigh-Jeans tail of
high-temperature thermal surface emission or nonthermal emission arising from 
their magnetospheres.  Optical (or near-IR) fluxes of magnetars
lie above the spectra extrapolated from their blackbody-like X-ray components
(e.g., \citealt{hvk00,mer13}), which excludes the former case for magnetars.  
For the latter, polarized emission is expected and actually is 
observed in optical emission from radio pulsars (e.g., \citealt{slo+09}). 
While the current emission models, such as
the polar cap, outer gap, two-pole caustic, and striped pulsar wind
(see the detailed discussion in \citealt{slo+09} and references therein),
which involve radiation mechanisms including curvature, synchrotron,
or inverse Compton scattering radiation,
can not fully explain the well-studied Crab pulsar case, 5--10\% phase-averaged
linear polarization is detected in optical emission from several close
or young radio pulsars 
(e.g., the Crab pulsar, \citealt{slo+09,mor+13};
B0540$-$69, \citealt{mpb87,lun+11};
the Vela pulsar, \citealt{mig+07,mms14};
B0656+14, \citealt{ker+03,mig+15};
B1509$-$58, \citealt{ws00}).
The polarization upper limit of 5.6\% we have obtained suggests, 
while marginally, a different emission mechanism from that considered
in radio pulsars and thus supports a class of their own for magnetars.

Considering the surface magnetic fields of $B\sim 10^{15}$~G for magnetars,
\citet{egl02} have suggested that optical emission from them
could be similar to radio emission from radio pulsars, i.e., due to
synchrotron radiation from electron/positron pairs. 
In this scenario, strong linear polarization should be seen,
since pulsars' radio emission is known to have the degrees of linear
polarization (phase-averaged) in a range of from 10\% to as high as 
100\% (e.g., \citealt{gl98,wj08,han+09}).
Additionally, \citet{bt07} have suggested
ion cyclotron emission or curvature emission by electron/positron pairs
as two possible mechanisms for magnetars' optical emission. 
For the both mechanisms, certain degree of linear polarization might be 
expected.  Our measurement suggests zero or low linear polarization
in optical emission, not supporting the scenarios. 

However, the propagation effects in a magnetosphere may cause a strong
depolarization of optical emission. The natural wave modes of
optical wave in the pulsar magnetosphere are usually two orthogonal
linearly polarized modes: the ordinary mode polarized in $\vec{k} - \vec{B}$
plane and the extraordinary mode perpendicular to that plane. The
adiabatic evolution condition of the two linear modes is (see details
in \citealt*{wlh10})
\begin{equation}
\Gamma_{\rm ad} = \left| \frac{\Delta k}{2\phi_B'}\right| \simeq 4\times
10^{-5}\eta_3\gamma_3B_5^3\nu_{15}^{-4}\gg 1.
\end{equation}
Here $\eta=N/N_{\rm GJ}$ is the multiplicity, $\eta_3=\eta/10^3$,
$\gamma_3=\gamma/10^3$ with $\gamma$ the Lorentz factor of the
streaming plasma, $B_5=B/10^5$G and $\nu_{15}=\nu/10^{15}$Hz are
the magnetic field strength and wave frequency. Note that we
reasonably suppose $\phi_B'\sim 1/r_{\rm lc}$ ($r_{\rm lc}$ is the
light cylinder radius). Assuming a dipole, the magnetic field
strength can be written as
\begin{equation}
B=B_{*15}(r/R_*)^{-3}\,{\rm G}=9.2 B_{*15}P_{\rm 10s}^{-3}(r/r_{\rm lc})^{-3}\,{\rm G},
\end{equation}
where $B_*$ is the surface magnetic field, $B_{*15}=B/10^{15}$\,G,
$R_*$ the neutron star radius, and we set 
$R_*=10$\,km, $P_{\rm 10s}=P/10$s. For $r\go0.015r_{\rm lc}(\eta_3\gamma_3\nu_{15}^{-4})^{1/9}B_{*15}^{1/3}P_{\rm 10s}^{-1}$, 
we have $|\Gamma_{\rm ad}|\ll1$. Thus if it is in the outer
magnetosphere where magnetar optical emission is generated, the mode
evolution would be usually non-adiabatic, implying that the polarization
direction almost does not change when propagating through the
magnetosphere.

Considering the emission of a relativistic particle accelerated
along the magnetic field direction, the initial polarization state
should be the pure ordinary mode ($\vec{E}\parallel \vec{B}_\perp$) in
each direction of the $1/\gamma$ emission cone. The polarization
percentage of the combined wave of the cone obviously equals to
zero at the emission point. As pointed by \citet{cr79}, 
the polarization directions of the cone could be aligned by
adiabatic walking 
after propagating a distance, which gives a very high polarization
percentage such as in radio emission. However, if the optical emission
comes from the outer magnetosphere, adiabatic walking would not occur since
the mode evolution is non-adiabatic. The final polarization degree
should be close to zero in this case. In this emission model, if
the emission region extends to inner magnetosphere where mode
evolution is adiabatic, partial polarization are expected, which
may be the case of the Crab pulsar. For 4U~0142+61, the optical
emission region could be higher so that the mode evolution is
purely non-adiabatic everywhere, and the final polarization degree is
close to zero.

Even the initial emission at each height is highly polarized for
some other mechanisms, the total emission from the whole
magnetosphere could also be depolarized due to
the aberration/retardation effect. The observed emission at one phase
may come from different magnetic field curve planes where the local
polarization directions are different. Different from the radio band,
pulsar optical emission may extend from the inner magnetosphere to
light cylinder radius, such as the whole slot gap and outer gap
region, which makes the depolarization due to
the aberration/retardation effect much stronger. The detailed
depolarization degree depends on the emission geometry in
a magnetosphere. We also note that possible precession
of 4U~0142+61 with a period of $\sim$15 hours was reported \citep{mak+14},
although it was not confirmed by NuSTAR observations \citep{ten+15}. This
precession could cause the polarization degree to be further averaged down
in our 3 and 5 hours observations.

In addition to the pulsar magnetospheric origin considered above,
optical I-band emission from 4U~0142+61 could have 
other origin.  For example, part of it (the non-pulsed component) could 
arise from the debris disk around the magnetar. In a recently re-proposed 
model for magnetars (\citealt{mrr12}; following the pioneering work 
by \citealt{mor+88} and \citealt{pac90}),
they are suggested to be massive, highly magnetized white dwarfs with 
emission powered by their rotational energy (similar to pulsars).
The optical and IR SED of 4U~0142+61 then might be explained by 
the thermal surface emission from a massive white dwarf plus that 
from a surrounding disk (for details, see \citealt{rue+13}). 
For these possibilities, low linear polarization
(at most a few percent) would be expected, since 
the central star's emission can significantly lower the polarization level
in optical light from a debris disk system (although emission from
a debris disk may have high polarization; e.g., \citealt{gg15}), or
thermal surface emission from a star cannot be highly polarized
(e.g., \citealt{che+88}). Our constraint on the linear
polarization is actually consistent with these possibilities.

Our Subaru polarimetry likely represents the best effort for measuring
polarization in optical light from 4U~0142+61 or the known magnetars
in general, since they do not have detectable optical emission or are
extremely faint \citep{ok14}. However, it would be interesting to carry
out such observations when they, particularly 4U~0142+61, are in an outburst.
The comparison would provide further information for our understanding of
magnetars' emission mechanisms and related properties.
In the future, polarimetry with an extremely large telescope, such as 
the Thirty-Meter Telescope, would certainly help our understanding by
obtaining a much tight constraint or a measurement in a much shorter 
observation. The result would possibly
help determine the emission mechanism at optical bands for magnetars and 
even identify the true nature of magnetars.


\acknowledgements

This paper is based on data collected at Subaru Telescope, which is operated 
by the National Astronomical Observatory of Japan.

We thank the anonymous referee for helpful suggestions and
Takashi Hattori for taking the data for us.
This research was supported in part by the National Natural Science 
Foundation of China (11373055) and the Strategic Priority Research Program 
``The Emergence of Cosmological Structures" of the Chinese Academy 
of Sciences (Grant No. XDB09000000). C. Wang acknowledeges the support
from the National Natural Science Foundation of China (11273029).

{\it Facility:} \facility{Subaru (FOCAS)}

\bibliographystyle{apj}

\begin{thebibliography}{31}
\expandafter\ifx\csname natexlab\endcsname\relax\def\natexlab#1{#1}\fi

\bibitem[{{Beloborodov} \& {Thompson}(2007)}]{bt07}
{Beloborodov}, A.~M., \& {Thompson}, C. 2007, \apj, 657, 967

\bibitem[{{Cheng} \& {Ruderman}(1979)}]{cr79}
{Cheng}, A.~F., \& {Ruderman}, M.~A. 1979, \apj, 229, 348

\bibitem[{{Cheng} {et al.}(1988){Cheng}, {Shields}, {Lin}, D.~N.~C. \& {Pringle}}]{che+88}
{Cheng}, F.~H., {Shields}, G.~A., {Lin}, D.~N.~C., \& {Pringle}, J.~E. 1988, \apj, 328, 223


\bibitem[{{Durant} \& {van Kerkwijk}(2006{\natexlab{a}})}]{dv06a}
{Durant}, M., \& {van Kerkwijk}, M.~H. 2006{\natexlab{a}}, \apj, 650, 1070

\bibitem[{{Durant} \& {van Kerkwijk}(2006{\natexlab{b}})}]{dv06b}
---. 2006{\natexlab{b}}, \apj, 650, 1082

\bibitem[{{Eichler} {et~al.}(2002){Eichler}, {Gedalin}, \& {Lyubarsky}}]{egl02}
{Eichler}, D., {Gedalin}, M., \& {Lyubarsky}, Y. 2002, \apjl, 578, L121

\bibitem[{{Garc{\'{\i}}a} \& {G{\'o}mez}(2015)}]{gg15}
{Garc{\'{\i}}a}, L., \& {G{\'o}mez}, M. 2015, Rev. Mexicana Astron. Astrofis., 51, 3

\bibitem[{{Gonzalez} {et~al.}(2010){Gonzalez}, {Dib}, {Kaspi}, {Woods}, {Tam},
  \& {Gavriil}}]{gon+10}
{Gonzalez}, M.~E., {Dib}, R., {Kaspi}, V.~M., {Woods}, P.~M., {Tam}, C.~R., \&
  {Gavriil}, F.~P. 2010, \apj, 716, 1345

\bibitem[{{Gould} \& {Lyne}(1998)}]{gl98}
{Gould}, D.~M., \& {Lyne}, A.~G. 1998, \mnras, 301, 235

\bibitem[{{Han} {et~al.}(2009){Han}, {Demorest}, {van Straten}, \&
  {Lyne}}]{han+09}
{Han}, J.~L., {Demorest}, P.~B., {van Straten}, W., \& {Lyne}, A.~G. 2009,
  \apjs, 181, 557

\bibitem[{{Hulleman} {et~al.}(2000){Hulleman}, {van Kerkwijk}, \&
  {Kulkarni}}]{hvk00}
{Hulleman}, F., {van Kerkwijk}, M.~H., \& {Kulkarni}, S.~R. 2000, \nat, 408,
  689
  
\bibitem[Hulleman et 
al.(2004)]{Hulleman04} Hulleman, F., van Kerkwijk, M.~H., \& Kulkarni, S.~R.\ 2004, \aap, 416, 1037 

\bibitem[Hurley et al.(1999)]{Hurley99} Hurley, K., Cline, T., 
Mazets, E., et al.\ 1999, \nat, 397, 41 

\bibitem[{{Kashikawa} {et~al.}(2002){Kashikawa}, {Aoki}, {Asai}, {Ebizuka},
  {Inata}, {Iye}, {Kawabata}, {Kosugi}, {Ohyama}, {Okita}, {Ozawa}, {Saito},
  {Sasaki}, {Sekiguchi}, {Shimizu}, {Taguchi}, {Takata}, {Yadoumaru}, \&
  {Yoshida}}]{kas+02}
{Kashikawa}, N., {et~al.} 2002, \pasj, 54, 819

\bibitem[{{Kaspi}(2007)}]{kas07}
{Kaspi}, V.~M. 2007, \apss, 308, 1

\bibitem[{{Kawabata} {et~al.}(1999){Kawabata}, {Okazaki}, {Akitaya}, 
{Hirakata}, {Hirata}, {Ikeda}, {Kondoh}, {Masuda}, \& {Seki}}]{kaw+99}
{Kawabata}, K.~S., {Okazaki}, A., {Akitaya}, H., et al. 1999, PASP, 111, 898

\bibitem[{{Kern} \& {Martin}(2002)}]{km02}
{Kern}, B., \& {Martin}, C. 2002, \nat, 417, 527

\bibitem[{{Kern} {et~al.}(2003){Kern}, {Martin}, {Mazin}, \&
  {Halpern}}]{ker+03}
{Kern}, B., {Martin}, C., {Mazin}, B., \& {Halpern}, J.~P. 2003, \apj, 597,
  1049

\bibitem[{{Lundqvist} {et~al.}(2011){Lundqvist}, {Lundqvist}, {Bj{\"o}rnsson},
{Olofsson}, {Pires}, {Shibanov}, \& {Zyuzin}}]{lun+11}
{Lundqvist}, N., {Lundqvist}, P., {Bj{\"o}rnsson}, C.-I., {Olofsson}, G.,
{Pires}, S., {Shibanov}, Y.~A., \& {Zyuzin}, D.~A. 2011, \mnras, 413, 611


\bibitem[{{Makishima} {et~al.}(2014){Makishima}, {Enoto}, {Hiraga}, {Nakano}, {Nakazawa}, {Sakurai}, {Sasano}, \& {Murakami}}]{mak+14}
{Makishima}, K., {Enoto}, T., {Hiraga}, J.~S., et al. 2014, Physical Review Letters, 112, 171102

\bibitem[{{Malheiro} {et~al.}(2012){Malheiro}, {Rueda}, \& {Ruffini}}]{mrr12}
{Malheiro}, M., {Rueda}, J.~A., \& {Ruffini}, R. 2012, PASJ, 64, 56

\bibitem[{{Mereghetti}(2013)}]{mer13}
{Mereghetti}, S. 2013, Brazilian Journal of Physics, 43, 356

\bibitem[{{Middleditch} {et~al.}(1987){Middleditch}, {Pennypacker}, \&
  {Burns}}]{mpb87}
{Middleditch}, J., {Pennypacker}, C.~R., \& {Burns}, M.~S. 1987, \apj, 315, 142

\bibitem[{{Mignani} {et~al.}(2007){Mignani}, {Bagnulo}, {Dyks}, {Lo Curto}, \&
  {S{\l}owikowska}}]{mig+07}
{Mignani}, R.~P., {Bagnulo}, S., {Dyks}, J., {Lo Curto}, G., \&
  {S{\l}owikowska}, A. 2007, \aap, 467, 1157

\bibitem[{{Mignani} {et~al.}(2015)}]{mig+15}
{Mignani}, R.~P., {Moran}, P., {Shearer}, A, {Testa}, V., {Slowikowska}, A.,
{Rudak}, B., {Krzeszowki}, K., \& {Kanbach}, G. 2015, \aap, arXiv:1510.01057


\bibitem[{{Moran} {et~al.}(2013){Moran}, {Shearer}, {Mignani}, {S{\l}owikowska}, {De Luca}, {Gouiff{\`e}s}, \& {Laurent}}]{mor+13}
{Moran}, P., {Shearer}, A., {Mignani}, R.~P., {S{\l}owikowska}, A., {De Luca}, A., {Gouiff{\`e}s}, C., \& {Laurent}, P. 2013, MNRAS, 433, 2564

\bibitem[{{Moran} {et~al.}(2014){Moran}, {Mignani}, \& {Shearer}}]{mms14}
{Moran}, P., {Mignani}, R.~P., \& {Shearer}, A. 2014, \mnras, 445, 835

\bibitem[{{Morini} {et~al.}(1988){Morini}, {Robba}, {Smith}, \& {van der Klis}}]{mor+88}
{Morini}, M., {Robba}, N.~R., {Smith}, A., \& {van der Klis}, M. 1988, \apj, 333, 777

\bibitem[{{Olausen} \& {Kaspi}(2014)}]{ok14}
{Olausen}, S.~A., \& {Kaspi}, V.~M. 2014, \apjs, 212, 6

\bibitem[{{Paczynski} (1990)}]{pac90}
{Paczynski}, B 1990, \apjl, 365, L9

\bibitem[{{Rea} {et~al.}(2010){Rea}, {Esposito}, {Turolla}, {Israel}, {Zane},
  {Stella}, {Mereghetti}, {Tiengo}, {G{\"o}tz}, {G{\"o}{\u g}{\"u}{\c s}}, \&
  {Kouveliotou}}]{rea+10}
{Rea}, N., {et~al.} 2010, Science, 330, 944

\bibitem[{{Rea} {et~al.}(2012){Rea}, {Israel}, {Esposito}, {Pons},
  {Camero-Arranz}, {Mignani}, {Turolla}, {Zane}, {Burgay}, {Possenti},
  {Campana}, {Enoto}, {Gehrels}, {G{\"o}{\v g}{\"u}{\c s}}, {G{\"o}tz},
  {Kouveliotou}, {Makishima}, {Mereghetti}, {Oates}, {Palmer}, {Perna},
  {Stella}, \& {Tiengo}}]{rea+12}
---. 2012, \apj, 754, 27

\bibitem[{{Rueda} {et~al.}(2013)}]{rue+13}
{Rueda}, J.~A., {Boshkayev}, K., {Izzo}, L., et al. 2013, \apjl, 772, L24


\bibitem[{{Schmidt} {et~al.}(1992){Schmidt}, {Elston}, \& {Lupie}}]{sel92}
{Schmidt}, G.~D., {Elston}, R., \& {Lupie}, O.~L. 1992, \aj, 104, 1563

\bibitem[{{S{\l}owikowska} {et~al.}(2009){S{\l}owikowska}, {Kanbach}, {Kramer},
  \& {Stefanescu}}]{slo+09}
{S{\l}owikowska}, A., {Kanbach}, G., {Kramer}, M., \& {Stefanescu}, A. 2009,
  \mnras, 397, 103

\bibitem[Tanaka et al.(2007)]{Tanaka07} Tanaka, Y.~T., Terasawa, 
T., Kawai, N., et al.\ 2007, \apjl, 665, L55 

\bibitem[{{Tendulkar} {et al.}(2015)}]{ten+15}
{Tendulkar}, S.~P., {Hasc{\"o}et}, R., {Yang}, C., et al. 2015, \apj, 808, 32

\bibitem[Terasawa et al.(2005)]{Terasawa05} Terasawa, T., Tanaka, 
Y.~T., Takei, Y., et al.\ 2005, \nat, 434, 1110 


\bibitem[{{Wagner} \& {Seifert}(2000)}]{ws00}
{Wagner}, S.~J., \& {Seifert}, W. 2000, in Astronomical Society of the Pacific
  Conference Series, Vol. 202, IAU Colloq. 177: Pulsar Astronomy - 2000 and
  Beyond, ed. M.~{Kramer}, N.~{Wex}, \& R.~{Wielebinski}, 315

\bibitem[{{Wang} {et~al.}(2010){Wang}, {Lai}, \& {Han}}]{wlh10}
{Wang}, C., {Lai}, D., \& {Han}, J. 2010, \mnras, 403, 569

\bibitem[{{Wang}(2014)}]{wan14}
{Wang}, Z. 2014, \planss, 100, 19

\bibitem[{{Wang} {et~al.}(2006){Wang}, {Chakrabarty}, \& {Kaplan}}]{wck06}
{Wang}, Z., {Chakrabarty}, D., \& {Kaplan}, D.~L. 2006, \nat, 440, 772

\bibitem[{{Wang} {et~al.}(2008){Wang}, {Chakrabarty}, \& {Kaplan}}]{wck08}
{Wang}, Z., {Chakrabarty}, D., \& {Kaplan}, D.~L. 2008, in American Institute
  of Physics Conference Series, Vol. 983, 40 Years of Pulsars: Millisecond
  Pulsars, Magnetars and More, ed. C.~{Bassa}, Z.~{Wang}, A.~{Cumming}, \&
  V.~M. {Kaspi}, 274--276

\bibitem[{{Wang} {et~al.}(2012){Wang}, {Tanaka}, \& {Zhong}}]{wtz12}
{Wang}, Z., {Tanaka}, Y.~T., \& {Zhong}, J. 2012, \pasj, 64, L1

\bibitem[{{Weltevrede} \& {Johnston}(2008)}]{wj08}
{Weltevrede}, P. \& {Johnston}, S. 2008, \mnras, 391, 1210


\bibitem[{{Woods} \& {Thompson}(2006)}]{wt06}
{Woods}, P.~M., \& {Thompson}, C. 2006, {Soft gamma repeaters and anomalous
  X-ray pulsars: magnetar candidates}, ed. W.~H.~G. {Lewin} \& M.~{van der
  Klis}, 547--586

\end{thebibliography}

\clearpage


\begin{table}
\begin{center}
\caption{PSF fitting photometry of 4U~0142+61}
\label{tab:phot}
    \begin{tabular}{lccc}
\hline
Sub-image & $m_{\rm psf}$ & $\Delta m_{{\rm cor}}$ & $m_{r=10}$ \\ 
	  &  (mag)          &   (mag)     & (mag)           \\\hline
e-beam$_{0\fdg0}$ & 26.809$\pm$0.071 & 0.250 & 26.559$\pm$0.071 \\
o-beam$_{0\fdg0}$ & 26.858$\pm$0.072 & 0.246 & 26.612$\pm$0.072 \\
e-beam$_{45\fdg0}$ & 26.780$\pm$0.073 & 0.255 & 26.525$\pm$0.073 \\
o-beam$_{45\fdg0}$ & 26.792$\pm$0.078 & 0.256  & 26.536$\pm$0.078 \\ 
e-beam$_{22\fdg5}$ & 26.693$\pm$0.065 & 0.257 & 26.436$\pm$0.065 \\
o-beam$_{22\fdg5}$ & 26.846$\pm$0.081 & 0.253  & 26.593$\pm$0.081 \\ 
e-beam$_{67\fdg5}$ & 26.731$\pm$0.060 & 0.260 & 26.471$\pm$0.060 \\
o-beam$_{67\fdg5}$ & 26.874$\pm$0.076 & 0.255  & 26.619$\pm$0.076 \\ 
\hline
    \end{tabular}
\vskip 1mm
\footnotesize{Note: instrumental magnitude $m=28-2.5\log({\rm flux})$; 
$m_{\rm psf}$, $\Delta m_{{\rm cor}}$, and $m_{r=10}$ are magnitudes 
obtained from PSF fitting, aperture corrections  
to a radius of 10 pixels (the uncertainties are negligible), 
and corrected magnitudes.}
 \end{center}
\end{table}

\clearpage

\end{document}